\documentclass[journal]{IEEEtran}
\usepackage{textcomp}
\IEEEoverridecommandlockouts
\usepackage{cite}
\usepackage{amsmath,amssymb,amsfonts, amsmath, dsfont}
\usepackage{bbm}
\usepackage{graphicx}
\usepackage{textcomp}
\usepackage{xcolor}
\usepackage{multirow}
\usepackage{subcaption}
\usepackage{cite}
\usepackage{textcomp}
\usepackage{dirtytalk}
\usepackage{textcomp}
\usepackage{makecell}
\usepackage{siunitx}
\usepackage{tabularx}
\usepackage{array}
\usepackage{bm}
\usepackage{bbm}
\usepackage{orcidlink}
\hypersetup{hidelinks}

\usepackage[utf8]{inputenc}
\usepackage{booktabs}   
\usepackage{tcolorbox}  
\usepackage{amsmath}    
\usepackage{enumitem}

\usepackage{tikz}
\usepackage{xcolor, colortbl}
\usetikzlibrary{decorations.pathreplacing}
\usetikzlibrary{arrows,shapes,positioning}
\usetikzlibrary{patterns}
\usepackage{pgfplots}
\pgfplotsset{compat=newest}
\pgfplotsset{plot coordinates/math parser=false}
\pgfkeys{/pgf/number format/.cd,1000 sep={\,}}
\usetikzlibrary{plotmarks}

\newlength\matlabfigurewidth
\newcolumntype{L}[1]{>{\raggedright\let\newline\\\arraybackslash\hspace{0pt}}m{#1}}
\newcolumntype{C}[1]{>{\centering\let\newline\\\arraybackslash\hspace{0pt}}m{#1}}
\newcolumntype{R}[1]{>{\raggedleft\let\newline\\\arraybackslash\hspace{0pt}}m{#1}}

\usepackage{algorithm}
\usepackage{algpseudocode}
\usepackage{graphicx}

\usepackage{booktabs}
\usepackage{amssymb}
\usepackage{pifont}
\usepackage{array}

\usepackage{graphicx}
\usepackage{tikz}
\usepackage{pgfplots}
\usepackage{subcaption}
\pgfplotsset{compat=1.18}
\usepgfplotslibrary{groupplots}

\begin{document}


\title{
Robust Joint Planning of EV and eBus Charging Infrastructure with PV Self-Consumption under Demand Uncertainty
}

\author{\IEEEauthorblockN{Biswarup Mukherjee\usepackage{0000-0001-7553-4339}, Member IEEE}\\
\IEEEauthorblockA{\textit{{}}\\
}
}

\author{Biswarup Mukherjee$^{\orcidlink{0000-0001-7553-4339}}$,~\IEEEmembership{Member,~IEEE}%
\thanks{Corresponding author: Biswarup Mukherjee (E-mail: bismuk@ieee.org).}%
\thanks{}}

\maketitle

\begin{abstract}
This paper presents a mixed-integer linear programming (MILP) framework for joint electric vehicle (EV) and electric bus (eBus) charging-infrastructure planning with photovoltaic (PV) self-consumption. The model co-optimizes charger siting, sizing, technology selection, eBus-to-depot assignment, and hourly charging schedules. A deterministic MILP is first formulated as a nominal benchmark and then extended to a scenario-based robust min--max formulation under vehicle energy-demand uncertainty. The robust model uses shared first-stage infrastructure decisions and scenario-specific operating decisions, and minimizes infrastructure cost plus the worst-case scenario operating cost through an epigraph reformulation. 
Soft-feasibility penalties quantify unmet charging energy, terminal state-of-charge (SOC) shortfall, and capacity violations under stressed scenarios. 
A 50-node case study shows that route-segment eBus demand and PV weighting alter charger deployment, technology mix, and grid import, while V2G enables hard-feasible robust operation at sufficiently large penalty weights. These results highlight that effective long-term infrastructure planning must simultaneously account for diverse vehicle fleets, operational strategies that adapt to different scenarios, and node-level constraints on charging capacity.

\end{abstract}


\section{Introduction}

The rapid electrification of both private electric vehicles (EVs) and urban public electric bus (eBus) fleets has introduced unprecedented spatiotemporal stresses on electrical power distribution networks (EPDNs).
As a result, determining infrastructure deployment while balancing transportation operational requirements has emerged as a crucial area of engineering research.

To date, a major body of literature has explored charger siting, sizing, and network-coordinated planning exclusively tailored for private EV charging infrastructure~\cite{Kapoor2024EPDN, Wen2024RobustEV, Mukherjee2026TwoStageEV, Zhao2025EV, sun2023data}. Recent studies have made significant advancements in coupled transportation–distribution network deployment \cite{Kapoor2024EPDN, Chen2025RobustEV, Wen2024RobustEV}, optimization under dynamic grid conditions \cite{Alizadeh2025Optimal, Nguyen2023DRMPC}, and robust and uncertainty-aware facility allocation under multi-dimensional uncertainties \cite{Chen2025RobustEV, Wen2024RobustEV, deb2021robust}.
In parallel, extensive research has focused on eBus charging infrastructure planning, including charger location and sizing \cite{Teichert2019, Ferro2023EBus, Najafi2025IntegratedEBus, liu2021optimizing}, charging technology selection \cite{Teichert2019}, depot planning \cite{Najafi2025IntegratedEBus, Ferro2023EBus}, PV/storage-integrated planning \cite{liu2023optimal}, battery swapping \cite{Kocer2023BSS}, operational scheduling \cite{Whitaker2023Network, Kang2025RobustEBusV2G}, and robust charging network design \cite{Kang2025RobustEBusV2G, Loaiza2026RobustEBus}.
Despite these advances, private EV and eBus charging infrastructures are almost exclusively planned as independent systems, even though both compete for the same distribution-network capacity, integration of renewable energy resources, and charging investments.
Furthermore, infrastructure planning models typically determine charger siting, sizing, and technology selection while assuming fixed operational characteristics, such as predetermined depot assignments or simplified charging demand~\cite{mukherjee2022optimal}. Conversely, operational scheduling studies optimize charging decisions for an already deployed infrastructure but do not influence long-term investment decisions~\cite{Qi2025HDRLEBusCharging,Whitaker2023Network,Nguyen2023DRMPC, bara2025multi, knezovic2017active}.

This traditional paradigm neglects the deep mathematical coupling that binds long-term infrastructure investment to multi-fleet operational flexibility. In a combined ecosystem, high-level operational choices—such as the spatial routing of private EVs, the physical depot assignment of schedule-constrained eBuses, and the hourly coordination of battery charging states—directly dictate the minimum required investment in fast or slow charger configurations.
Consequently, infrastructure investment decisions should be directly coupled with operational variables—including depot assignment, charger utilization, vehicle state-of-charge (SOC) evolution, and hourly charging schedules—to accurately capture long-term infrastructure requirements.
A failure to co-optimize these dimensions yields infrastructure designs that are either excessively over-engineered (inflating capital expenditure) or chronically vulnerable to grid capacity violations during peak demand intervals.
More broadly, smart-charging and grid-integration studies have emphasized that charging infrastructure, charging schedules, distribution-grid capacity, and renewable-energy utilization should be coordinated to ensure reliable and economically efficient system operation~\cite{IRENA2019SmartCharging,Rather2021EVGridIndia, mukherjee2023optimized, mukherjee2023optimization}.

Recent studies have increasingly adopted robust optimization frameworks to hedge against uncertain charging demand, renewable generation, electricity prices, transportation demand, and infrastructure failures~\cite{Wen2024RobustEV,Chen2025RobustEV,Kang2025RobustEBusV2G,Mukherjee2026TwoStageEV}. These studies significantly improve infrastructure reliability under uncertainty; however, they remain focused on either private EV charging systems or public eBus fleets individually.
Furthermore, planning-operation integration is generally limited to homogeneous transportation systems in which infrastructure investment decisions are optimized together with a single class of charging demand.
To the best of the authors' knowledge, a unified planning framework that simultaneously considers heterogeneous EV and eBus charging demand, shared distribution-network resources, depot assignment, PV self-consumption, operational charging schedules, battery dynamics, and robust demand uncertainty has not yet been reported.

Despite the rapid progress in EV charging-infrastructure planning, eBus charging optimization, and robust planning under uncertainty, three important gaps remain. 
First, existing robust planning frameworks generally focus on either private EVs or eBuses and therefore do not capture the coupling that arises when heterogeneous fleets compete for common distribution-network capacity, local renewable resources, and infrastructure investment. 
Second, planning and operation are often only partially integrated, with long-term infrastructure decisions determined separately from operational decisions such as depot assignment and charging schedules. 
Third, simultaneously representing multi-fleet infrastructure allocation, depot assignment, vehicle-level charging schedules, 
and scenario-dependent recourse can lead to large and computationally demanding optimization problems for urban 
planners. 
These gaps motivate a unified and tractable formulation that co-optimizes joint EV/eBus charging infrastructure, depot assignment, operational charging schedules, PV self-consumption, and transportation-demand uncertainty within a MILP framework.
The primary contributions of this work are summarized as follows:
\begin{itemize}
\item \textit{Heterogeneous multi-fleet co-optimization framework:} A unified framework is proposed to simultaneously addresses the planning and operational scheduling of charging infrastructure for private EVs and eBuses within a shared distribution network. Unlike approaches that treat the two fleets separately, the proposed model co-optimizes charger siting, sizing, technology selection, eBus-to-depot assignment, and hourly EV/eBus charging while explicitly incorporating common distribution-grid hosting limits and the available local PV resources.

\item \textit{Tractable two-stage robust reformulation:} 
A scenario-based two-stage robust MILP is developed to hedge against EV and eBus demand uncertainty. Charger deployment and eBus-to-depot assignment are represented as shared first-stage decisions, whereas charging schedules, SOC trajectories, grid import, PV utilization, and soft-feasibility variables constitute scenario-dependent operational recourse. An epigraph reformulation minimizes infrastructure investment together with the worst-case scenario operating cost.

\item \textit{Quantification of capital-robustness trade-offs:} The framework enables quantitative evaluation of the trade-offs among infrastructure investment, PV self-consumption, 
grid energy import, operational feasibility, and robustness under uncertain transportation demand, providing practical planning insights unavailable from deterministic planning approaches.
\end{itemize}
The remainder of this paper is organized as follows. Section~II presents the common physical model and notation. Section~III formulates the centralized deterministic and two-stage robust MILPs. Section~IV describes the case-study configuration and implementation details. Section~V presents and discusses the numerical results. Finally, Section~VI concludes the paper.

\begin{table*}[t]
\centering
\small
\caption{Taxonomic comparison of the proposed framework against existing literature.}
\label{tab:taxonomy_compact}
\renewcommand{\arraystretch}{1.3}
\begin{tabularx}{\textwidth}{@{} >{\raggedright\arraybackslash}p{2.6cm} @{\hspace{4pt}} c c @{\hspace{10pt}} >{\raggedright\arraybackslash}X @{\hspace{12pt}} >{\raggedright\arraybackslash}X @{\hspace{12pt}} >{\raggedright\arraybackslash}X @{\hspace{10pt}} c c c @{}}
\toprule
\makecell[b]{\textbf{Literature /}\\\textbf{references}} & 
\makecell[b]{\textbf{Target}\\\textbf{fleet}} & 
\makecell[b]{\textbf{Joint EV+}\\\textbf{eBus}} & 
\makecell[b]{\textbf{Planning}\\\textbf{scope}} & 
\makecell[b]{\textbf{Operational}\\\textbf{scheduling}} & 
\makecell[b]{\textbf{Modeling}} & 
\makecell[b]{\textbf{PV}\\\textbf{integ.}} & 
\makecell[b]{\textbf{Depot}\\\textbf{assign.}} & 
\makecell[b]{\textbf{Shared}\\\textbf{res.}} \\
\midrule

\makecell[l]{\cite{Kapoor2024EPDN}, \cite{Mukherjee2026TwoStageEV}, \cite{sun2023data},\\ \cite{Alizadeh2025Optimal}, \cite{mukherjee2022optimal}, \cite{He2023Coordinated}}
& Private EV & -- & Siting, sizing \& port selection & Static / Real-time flow & Deterministic / dynamic & -- & -- & -- \\
\addlinespace

\makecell[l]{\cite{Wen2024RobustEV}, \cite{Zhao2025EV}, \cite{Chen2025RobustEV},\\ \cite{Nguyen2023DRMPC}, \cite{deb2021robust}}
& Private EV & -- & Siting \& resilient~allocation & MPC / network flows & Robust / DRO / chaos & Partial \cite{Nguyen2023DRMPC} & -- & -- \\
\addlinespace

\makecell[l]{\cite{Teichert2019}, \cite{Ferro2023EBus}, \cite{Najafi2025IntegratedEBus},\\ \cite{liu2021optimizing}, \cite{Kocer2023BSS}, \cite{Whitaker2023Network}}
& Public eBus & -- & Siting, sizing \& swapping & Timetabling / network flow & Deterministic & -- & \checkmark & -- \\
\addlinespace

\makecell[l]{\cite{liu2023optimal}, \cite{Kang2025RobustEBusV2G}, \cite{Loaiza2026RobustEBus},\\ \cite{Qi2025HDRLEBusCharging}}
& Public eBus & -- & Battery / charger sizing & V2G rules / DRL & Robust optimization & \checkmark \cite{liu2023optimal} & \checkmark & -- \\
\addlinespace

\makecell[l]{\cite{bara2025multi}, \cite{IRENA2019SmartCharging}, \cite{Rather2021EVGridIndia},\\ \cite{mukherjee2023optimized}}
& Private EV & -- & Smart charging \& PV self-cons. & Hourly dispatch / control & Deterministic / dynamic & \checkmark \cite{mukherjee2023optimized} & -- & Partial \\
\midrule

\textbf{This paper} & \textbf{EV \& eBus} & \textbf{\checkmark} & \textbf{Siting, sizing \& technology selection} & \textbf{Hourly charging co-optimization} & \textbf{Two-stage robust MILP} & \textbf{\checkmark} & \textbf{\checkmark} & \textbf{\checkmark} \\
\bottomrule
\end{tabularx}
\end{table*}

\section{Methodology}
This section presents the MILP formulation for joint EV and eBus charging infrastructure planning with PV self-consumption. The model determines the number, type, and location of chargers, the assignment of eBuses to depots, and the hourly charging schedules of EVs and eBuses over a finite time horizon. Two formulations are considered: (i) a centralized deterministic MILP, which serves as the nominal benchmark, and (ii) a scenario-based robust min--max MILP, in which charger deployment decisions are shared across demand scenarios while operational decisions are scenario-specific.

\subsection{Overall framework}
\begin{figure}[t!]
    \centering
    \includegraphics[width=0.9\columnwidth]{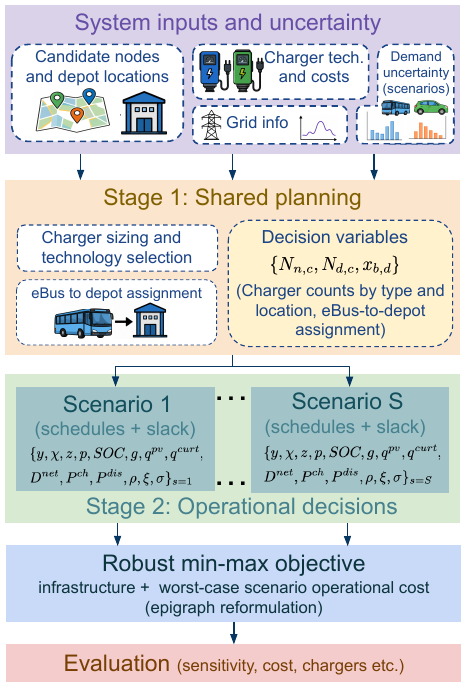}
    \caption{Overall methodology of the proposed robust joint EV and eBus charging-infrastructure planning framework.
    }
    \label{fig:overall_method}
\end{figure}

Figure~\ref{fig:overall_method} summarizes the proposed framework. Inputs include vehicle availability and energy-demand profiles, candidate charging and depot locations, charger characteristics, PV availability, grid-import price, and nodal hosting limits. Demand uncertainty is represented by a finite scenario set: EV scenarios modify daily charging-energy requirements and terminal SOC targets, whereas eBus scenarios scale service-energy consumption in the SOC dynamics.

The deterministic MILP provides the nominal benchmark. 
In the robust formulation, planning decisions are shared across demand scenarios, while operational decisions are treated as scenario-dependent recourse. 
The resulting two-stage model is solved as a single integrated min--max MILP using an epigraph representation of the worst-case operating cost.

\subsection{Sets, indices, and notation}

The principal sets are the candidate nodes $\mathcal{N}$, eBus depots $\mathcal{D} \subseteq \mathcal{N}$, time periods $\mathcal{T}$, EVs $\mathcal{V}$, eBuses $\mathcal{B}$, scenarios $\mathcal{S}$, and charger types $\mathcal{C}^{\mathrm{ev}}$ and $\mathcal{C}^{\mathrm{bus}}$. Tables~\ref{tab:parameters} and \ref{tab:variables} summarize the complete mathematical notation used throughout the paper. 
In Table~\ref{tab:variables}, the scenario index is omitted for operational variables whose deterministic and robust versions share the same physical definition; the robust formulation reintroduces the index $s$ where scenario dependence is explicit.

\begin{table}[htb!]
\caption{Sets, indices, and parameters}
\label{tab:parameters}
\footnotesize
\centering
\setlength{\tabcolsep}{3pt}
\renewcommand{\arraystretch}{0.9}
\begin{tabularx}{\columnwidth}
{>{\raggedright\arraybackslash}p{0.3\columnwidth}
>{\raggedright\arraybackslash}X}
\hline
\textbf{Symbol} & \textbf{Description}\\
\hline

$\mathcal{N},\mathcal{D},\mathcal{T}$
& Candidate charging nodes, eBus depot nodes
($\mathcal{D}\subseteq\mathcal{N}$), and time periods, respectively. \\


$\mathcal{V},\mathcal{B},\mathcal{S}$
& Sets of EVs, eBuses, and demand scenarios, respectively. \\

$\mathcal C^{ev},\mathcal C^{bus}$ & Sets of EV and eBus charger types.\\
$\mathcal C_n$ & Charger types installable at node $n$.\\

$n,t,s,v,b,d,r,c$
& Node, time, scenario, EV, eBus, depot, route-segment, and charger-type indices. \\

$A^{dep}_{b,d,t}$ & eBus depot-availability indicator.\\

$\bar {P}^{f},\bar {P}^{s},\bar P^{\mathrm{bus}}$
& Rated powers of fast EV, slow EV, and eBus charger ports,
respectively. \\

$\bar Q^{pv}_{n,t}$ & Available PV generation.\\
$\bar P^{node}_{n}$
&
Active-power charging-hosting limit at node $n$.
\\
$E^{ev},E^{bus}$ & Battery capacities of EVs and eBuses.\\
$E^{\mathrm{req}}_v,\;E^{\mathrm{req}}_b$
& Nominal EV charging-energy requirement and eBus minimum daily
charging-energy delivery floor, respectively. \\
$e^{\mathrm{serv}}_{b,t},
 e^{\mathrm{serv}}_{b,t,s}$
& Nominal and scenario-dependent eBus service-energy consumption
during period $t$. \\

$\mu_s$
& Transportation-energy multiplier associated with scenario $s$. \\

$E^{\mathrm{req}}_{v,s}$
& Scenario-dependent daily charging-energy requirement of EV $v$. \\
$\mathrm{SOC}^{\mathrm{ev},0}_{v},
 \mathrm{SOC}^{\mathrm{bus},0}_{b}$
& Initial SOC of EV $v$ and eBus $b$,
respectively. \\

$\overline{\mathrm{SOC}}^{\mathrm{ev}}_{v},
 \overline{\mathrm{SOC}}^{\mathrm{bus}}_{b}$
& Nominal terminal SOC targets of EV $v$ and eBus $b$, respectively. \\

$\overline{\mathrm{SOC}}^{\mathrm{ev}}_{v,s}$
& Scenario-dependent terminal SOC target of EV $v$. \\

$\eta^{ev},\eta^{bus}$ & Charging/discharging efficiencies.\\

$R_b, L_{b,r,t}$ & Set of route segments traversed by eBus $b$ and distance travelled on segment $r$.\\
$\alpha_r, \kappa_c$ & Route energy-consumption coefficient (kWh/km) and investment cost of charger type $c$ (k\$).\\

$\lambda,\omega$
& Grid-import price and PV self-consumption weighting coefficient (k\$/kWh),
respectively. \\

$M,\gamma^{\mathrm{v2g}}$
& Soft-feasibility penalty and battery-throughput cost (k\$/kWh) coefficients,
respectively. \\

$\mathbb{I}_{n\in\mathcal{D}},\delta_{\mathrm{V2G}}$
& Depot-node and V2G-enabled indicators, respectively. \\

\hline
\end{tabularx}
\end{table}

\begin{table}[t!]
\caption{Decision variables and derived quantities}
\label{tab:variables}
\footnotesize
\centering
\setlength{\tabcolsep}{3pt}
\renewcommand{\arraystretch}{0.95}
\begin{tabularx}{\columnwidth}
{>{\raggedright\arraybackslash}p{0.30\columnwidth}
>{\raggedright\arraybackslash}X}
\hline
\textbf{Symbol} & \textbf{Description}\\
\hline

$N_{n,c},N_{d,c}$
& Numbers of EV and eBus chargers installed at node $n$
and depot $d$, respectively. \\

$x_{b,d}$ & Binary eBus-to-depot assignment variable.\\
$y_{b,t,d}$ & eBus plug-in indicator.\\

$\chi^{f}_{v,t},\chi^{s}_{v,t}$ & EV fast/slow charger occupancy variables.\\

$z^{r,+}_{v,t},z^{r,-}_{v,t}$ & EV charging/discharging mode variables.\\
$z^{bus,+}_{b,t,d},z^{bus,-}_{b,t,d}$ & eBus charging/discharging mode variables.\\

$p^{ev,+}_{v,t},p^{ev,-}_{v,t}$ & EV charging/discharging power.\\
$p^{bus,+}_{b,t,d},p^{bus,-}_{b,t,d}$ & eBus charging/discharging power.\\

$SOC^{ev}_{v,t},SOC^{bus}_{b,t}$ & State of charge.\\

$g_{n,t},q^{\mathrm{pv}}_{n,t},q^{\mathrm{curt}}_{n,t}$
& Grid-import, PV self-consumption, and PV-curtailment power,
respectively. \\

$D^{net}_{n,t}$ & Net charging demand after V2G injection.\\
$P^{ch}_{n,t,s}, P^{dis}_{n,t,s}$ & Aggregate charging and discharging power, respectively.\\
$\rho^{ch}_{n,t,s}$
&
Installed charger-capacity slack in scenario $s$.
\\

$\rho^{grid}_{n,t,s}$
&
Nodal active-power hosting-capacity slack in scenario $s$.
\\
$\xi^{ev}_{v},\xi^{bus}_{b}$ & Unmet charging-energy slack variables.\\
$\sigma^{ev}_{v},\sigma^{bus}_{b}$ & Terminal SOC slack variables.\\

$\theta$ & Worst-case operating-cost epigraph variable.\\

$C^{\mathrm{cap}}$, $C^{\mathrm{grid}}$
& Infrastructure investment cost and deterministic grid-import operating cost. \\

$C^{\mathrm{grid}}_s$, $C^{\mathrm{pv}}_s$
& Scenario-$s$ grid-import operating cost and PV self-consumption benefit. \\

$C^{\mathrm{slack}}_s$, $C^{\mathrm{v2g}}_s$
& Scenario-$s$ soft-feasibility penalty and battery-throughput cost. \\

$C^{\mathrm{op}}_s$, $C^{\mathrm{real}}_s$
& Scenario-$s$ operating cost and physical system cost excluding soft-feasibility penalties. \\

$C^{\mathrm{pen}}_s$, $C^{\mathrm{rob}}$
& Scenario-$s$ penalized system cost and worst-case penalized robust solution cost. \\

$\Omega,\Omega^{}_{s}$ & Deterministic and robust violation measures.\\

\hline
\end{tabularx}
\end{table}

\subsection{Common physical constraints}

The deterministic and robust models share the following physical constraints. They are written without a scenario index for compactness; in the robust formulation (Section~\ref{sec:robust_formulation}), all operational variables and uncertain demand parameters are replicated for each $s\in\mathcal{S}$, while infrastructure and depot-assignment decisions remain shared across scenarios.

\subsubsection{eBus depot assignment and infrastructure requirements}

Assuming the binary variable $x_{b,d}$ is a planning decision equal to one if eBus $b$ is assigned to depot $d$, each eBus must be assigned to exactly one depot:
\begin{equation}
\label{eq:bus_assignment}
\sum_{d\in\mathcal{D}} x_{b,d} = 1,
\quad \forall b\in\mathcal{B}.
\end{equation}
We denote $A^{dep}_{b,d,t}\in\{0,1\}$ the depot-return availability parameter, which is equal to one if eBus $b$ is physically available at depot $d$ during time period $t$, and zero otherwise. The plug-in status $y_{b,t,d}$ of bus $b$ at depot $d$ is coupled with both depot assignment and depot availability:
\begin{equation}
\label{eq:bus_depot_availability}
y_{b,t,d}
\leq
x_{b,d} A^{dep}_{b,d,t},
\quad \forall b,t,d .
\end{equation}

To capture bidirectional eBus charging, $p^{bus,+}_{b,t,d}$ and $p^{bus,-}_{b,t,d}$ denote, respectively, the charging and discharging power of eBus $b$ at depot $d$ and time $t$. The binary variables $z^{bus,+}_{b,t,d}$ and $z^{bus,-}_{b,t,d}$ indicate whether the eBus is charging or discharging, respectively. Since simultaneous charging and discharging are not allowed, the operating mode is linked to the plug-in status as:
\begin{equation}
\label{eq:bus_charge_discharge_exclusive}
z^{bus,+}_{b,t,d}
+
z^{bus,-}_{b,t,d}
\leq
y_{b,t,d},
\quad \forall b,t,d .
\end{equation}
The corresponding charging and discharging powers are bounded by the selected operating mode as:
\begin{align}
\label{eq:bus_charge_couple}
p^{bus,+}_{b,t,d}
&\leq
\bar{P}^{bus} z^{bus,+}_{b,t,d},
&& \forall b,t,d, \\
\label{eq:bus_discharge_couple}
p^{bus,-}_{b,t,d}
&\leq
\bar{P}^{bus} z^{bus,-}_{b,t,d},
&& \forall b,t,d .
\end{align}
The number of simultaneously plugged eBuses and the aggregate bidirectional charging and discharging power available at depot $d$ are limited by the installed charger ports and their rated power. Mathematically this is expressed as:
\begin{align}
\label{eq:bus_port_limit}
\sum_{b\in\mathcal{B}} y_{b,t,d} &\leq N_{d,bs1} + 2N_{d,bm}, && \forall d,t, \\
\label{eq:bus_charge_capacity}
\sum_{b\in\mathcal{B}} p^{bus,+}_{b,t,d}
&\leq
\bar P^{\mathrm{bus}}(N_{d,bs1} + 2N_{d,bm}), && \forall d,t, \\
\label{eq:bus_discharge_capacity}
\sum_{b\in\mathcal{B}} p^{bus,-}_{b,t,d}
&\leq
\bar P^{\mathrm{bus}}(N_{d,bs1}
+
2N_{d,bm}),
&& \forall d,t .
\end{align}
In \eqref{eq:bus_port_limit}--\eqref{eq:bus_discharge_capacity}, $bs1$ and $bm$ denote single-port and two-port multi-port eBus chargers, respectively. 
These constraints capture the depot’s operational capacity by limiting the number of simultaneously connected buses and the total bidirectional power supported by the heterogeneous charging infrastructure.

\subsubsection{EV charging, discharging and infrastructure Constraints}

For each EV, binary variables $\chi^f_{v,t}$ and $\chi^s_{v,t}$ indicate whether EV $v$ occupies a fast or slow charging port at time period $t$, respectively. Binary variables $z^{f,+}_{v,t}$, $z^{s,+}_{v,t}$, $z^{f,-}_{v,t}$, and $z^{s,-}_{v,t}$ indicate charging and discharging through fast and slow chargers, respectively. The mutually exclusive port-occupancy and bidirectional charging logic are imposed as:
\begin{align}
\label{eq:ev_port_exclusive}
\chi^f_{v,t}+\chi^s_{v,t}
&\leq 1,
&& \forall v,t, \\
\label{eq:ev_charge_requires_port}
z^{r,+}_{v,t}+z^{r,-}_{v,t}
&\leq
\chi^r_{v,t},
&& \forall v,t,\; r\in\{f,s\}.
\end{align}
The charging and discharging powers are bounded by the selected charger type:
\begin{align}
\label{eq:ev_charge_power}
p^{ev,+}_{v,t}
&\leq
\bar P^f z^{f,+}_{v,t}
+
\bar P^s z^{s,+}_{v,t},
&& \forall v,t,
\\
\label{eq:ev_discharge_power}
p^{ev,-}_{v,t}
&\leq
\bar P^f z^{f,-}_{v,t}
+
\bar P^s z^{s,-}_{v,t},
&& \forall v,t.
\end{align}
Constraints~\eqref{eq:ev_port_exclusive}--\eqref{eq:ev_discharge_power} jointly enforce exclusive fast/slow port occupancy, mutually exclusive charging and discharging through the occupied port, and charger-rated
power limits. Outside valid parking intervals, the corresponding occupancy, operating-mode, and power variables are fixed to zero.

The number of occupied fast and slow EV charging ports is limited by the installed charger-port capacity:
\begin{align}
\label{eq:ev_fast_ports}
\sum_{v\in\mathcal{V}_{n,t}} \chi^f_{v,t}
&\leq
N_{n,f1}+4N_{n,fm},
&& \forall n,t, \\
\label{eq:ev_slow_ports}
\sum_{v\in\mathcal{V}_{n,t}} \chi^s_{v,t}
&\leq
N_{n,s1}+4N_{n,sm},
&& \forall n,t .
\end{align}
The aggregate EV charging and discharging powers at each node are limited by the installed heterogeneous EV charging infrastructure:
\begin{align}
\label{eq:ev_node_charge_power}
\sum_{v\in\mathcal{V}_{n,t}} p^{ev,+}_{v,t}
&\leq
\bar{P}^{f}
\left(
N_{n,f1}+4N_{n,fm}
\right)
\nonumber\\
&\quad+
\bar{P}^{s}
\left(
N_{n,s1}+4N_{n,sm}
\right),
&& \forall n,t,
\\
\label{eq:ev_node_discharge_power}
\sum_{v\in\mathcal{V}_{n,t}} p^{ev,-}_{v,t}
&\leq
\bar{P}^{f}
\left(
N_{n,f1}+4N_{n,fm}
\right)
\nonumber\\
&\quad+
\bar{P}^{s}
\left(
N_{n,s1}+4N_{n,sm}
\right),
&& \forall n,t.
\end{align}
In \eqref{eq:ev_fast_ports}--\eqref{eq:ev_node_discharge_power},
$f1$, $fm$, $s1$, and $sm$ denote fast single-port, fast multi-port, slow single-port, and slow multi-port EV chargers, respectively; each multi-port EV charger provides four ports.
%


\subsubsection{SOC dynamics model}

The SOC dynamics are enforced for all eBuses $b \in \mathcal{B}$, EVs $v \in \mathcal{V}$, and time periods $t \in \mathcal{T} \setminus \{T\}$. V2G operation is represented by modeling the charging and discharging power of eBuses at depots $d$ and of EVs at candidate nodes.
Specifically for eBuses, the traction energy consumption during operation is characterized using detailed route-segment information as:
\begin{equation}
\label{eq:route_service_energy}
e^{serv}_{b,t}
=
\sum_{r\in\mathcal{R}_b}
\alpha_{r} L_{b,r,t},
\quad \forall b,t,
\end{equation}
where $\mathcal{R}_b$ is the set of route segments served by bus $b$, $L_{b,r,t}$ is the distance traveled by bus $b$ on route segment $r$ during time period $t$, and $\alpha_r$ is the route-segment energy-consumption coefficient in kWh/km.

Thus, the inter-temporal battery SOC dynamics of eBus and EV can be written as:
\begin{align}
\label{eq:bus_soc_v2g}
\mathrm{SOC}^{bus}_{b,t+1}
&=
\mathrm{SOC}^{bus}_{b,t}
\nonumber\\
&\quad+
\frac{1}{E^{bus}}
\left(
\begin{aligned}
&
\eta^{bus}
\sum_{d\in\mathcal{D}}
p^{bus,+}_{b,t,d}\cdot\Delta t
\\
&
-
\frac{1}{\eta^{bus}}
\sum_{d\in\mathcal{D}}
p^{bus,-}_{b,t,d}\cdot\Delta t
\\
&
-
e^{serv}_{b,t}
\end{aligned}
\right),
\\
\label{eq:ev_soc_v2g}
\mathrm{SOC}^{ev}_{v,t+1}
&=
\mathrm{SOC}^{ev}_{v,t}
\nonumber\\
&\quad+
\frac{1}{E^{ev}}
\left(
\eta^{ev}p^{ev,+}_{v,t}
-
\frac{1}{\eta^{ev}}
p^{ev,-}_{v,t}
\right)\cdot \Delta t.
\end{align}
The initial battery states are fixed as
$\mathrm{SOC}^{ev}_{v,0}=\mathrm{SOC}^{ev,0}_{v}$ and
$\mathrm{SOC}^{bus}_{b,0}=\mathrm{SOC}^{bus,0}_{b}$.
In addition, the terminal SOC requirements are imposed with non-negative slack variables:
\begin{align}
\label{eq:ev_soc_final_soft}
\mathrm{SOC}^{ev}_{v,T}
+
\sigma^{ev}_{v}
&\geq
\overline{\mathrm{SOC}}^{ev}_{v},
&& \forall v, \\
\label{eq:bus_soc_final_soft}
\mathrm{SOC}^{bus}_{b,T}
+
\sigma^{bus}_{b}
&\geq
\overline{\mathrm{SOC}}^{bus}_{b},
&& \forall b, \\
\label{eq:soc_slack_nonnegative}
\sigma^{ev}_{v},\sigma^{bus}_{b}
&\geq 0,
&& \forall v,b .
\end{align}
The nonnegative terminal-SOC slacks preserve feasibility when the  targets cannot be fully reached and are converted to equivalent energy in the objective. 
Throughout the scheduling horizon, SOC is bounded within $[0.05,1]$ for EVs and $[0.10,0.95]$ for eBuses. The initial SOC values and terminal SOC targets are specified separately in the case-study setup.

\subsubsection{Energy delivery requirements}

In addition to terminal SOC requirements, the total delivered charging energy must satisfy the daily energy requirement of each vehicle. These constraints are modeled as:
\begin{align}
\label{eq:ev_energy_req_soft}
\sum_{t\in\mathcal T} \eta^{ev} p^{ev,+}_{v,t}\Delta t
+ \xi^{ev}_{v}
&\ge E^{req}_{v}, && \forall v\in\mathcal V, \\
\label{eq:bus_energy_req_soft}
\sum_{t\in\mathcal T}\sum_{d\in\mathcal D}
\eta^{bus} p^{bus,+}_{b,t,d}\Delta t
+ \xi^{bus}_{b}
&\ge E^{req}_{b}, && \forall b\in\mathcal B. \\
\label{eq:energy_slack_nonnegative}
\xi^{ev}_{v},\xi^{bus}_{b}
&\geq 0,
&& \forall v,b .
\end{align}
For eBuses, $E^{\mathrm{req}}_b$ is a minimum delivered-energy constraint and is distinct from the traction-energy term $e^{\mathrm{serv}}_{b,t}$ appearing in the SOC dynamics.
Variables $\xi^{ev}_{v}$ and $\xi^{bus}_{b}$ denote unmet EV and eBus charging energy, respectively. 
Any remaining unmet energy is penalized via the soft-feasibility term in the objective function.

\subsubsection{PV self-consumption and grid import}

At each node and time period, the local net charging demand is supplied by locally self-consumed PV generation and grid import as in \cite{mukherjee2023optimized}. Since both eBuses and EVs are allowed to operate in charging and discharging modes, the net power demand at node $n$ and time period $t$ is expressed as:
\begin{equation}
\label{eq:local_charging_demand}
\begin{split}
D^{net}_{n,t}
=&
\sum_{v\in\mathcal{V}_{n,t}}
\left(
p^{ev,+}_{v,t}
-
p^{ev,-}_{v,t}
\right) \\
&+
\mathbb{I}_{n\in\mathcal{D}}
\sum_{b\in\mathcal{B}}
\left(
p^{bus,+}_{b,t,n}
-
p^{bus,-}_{b,t,n}
\right),
\quad \forall n,t .
\end{split}
\end{equation}
Equation \eqref{eq:local_charging_demand} defines the net power demand by accounting for both charging and discharging powers of eBuses and EVs. Charging power contributes positively to the node demand, whereas V2G discharging offsets the local demand through power injection. The indicator $\mathbb{I}_{n\in\mathcal{D}}$ ensures that eBus charging and discharging are considered only at depot nodes.

The nodal power balance can be expressed as:
\begin{equation}
\label{eq:grid_pv_balance}
D^{net}_{n,t}
=
q^{pv}_{n,t}
+
g_{n,t},
\quad \forall n,t ,
\end{equation}
where $q^{pv}_{n,t}$ denotes the locally self-consumed PV power and $g_{n,t}$ denotes the grid import required after accounting for PV generation and V2G operation.
The PV self-consumption and curtailment constraints are:
\begin{align}
\label{eq:pv_limit}
0 \le q^{pv}_{n,t}
&\le
\bar{Q}^{pv}_{n,t},
&& \forall n,t,\\
\label{eq:pv_curtailment}
q^{curt}_{n,t}
&=
\bar{Q}^{pv}_{n,t}
-
q^{pv}_{n,t},
&& \forall n,t,\\
\label{eq:grid_nonnegative}
g_{n,t},\;
q^{curt}_{n,t}
&\ge0,
&& \forall n,t.
\end{align}

Equations~\eqref{eq:pv_limit}--\eqref{eq:grid_nonnegative}
impose behind-the-meter PV self-consumption without grid export. Because both $g_{n,t}$ and $q^{\mathrm{pv}}_{n,t}$ are nonnegative, V2G discharge can offset local charging demand but cannot produce net export to the grid; any PV generation that cannot be locally absorbed is curtailed.

\subsubsection{Nodal active-power hosting constraints}

Each grid node is subject to an active-power charging-hosting limit $\bar{P}_n^{\text{node}}$,
which restricts the transportation-related net demand defined in~\eqref{eq:grid_pv_balance}. The hosting-limit representation is an aggregate network abstraction; conventional non-transport demand is not introduced as a separate time-varying term in the current model.
The deterministic nodal active charging-hosting constraint is:
\begin{equation}
\label{eq:nodal_hosting_det}
D^{net}_{n,t}
\le
\bar P^{node}_{n},
\qquad
\forall n\in\mathcal N,\;t\in\mathcal T,
\end{equation}
In the bidirectional (V2G) case, discharge at grid node reduces the net power demand $D^{\mathrm{net}}_{n,t}$.

\section{The joint optimization model}

Based on the physical constraints and system modeling defined in Section~II, this section presents the complete joint optimization framework. First, the common objective function components are established, followed by the centralized deterministic formulation and its scenario-based robust extension under transportation-energy uncertainty.

\subsection{Common objective function components}

Both optimization formulations are constructed from the same set of economic cost components. These components quantify the infrastructure investment, operating cost, PV self-consumption benefit, and residual soft-feasibility penalty. Defining these terms separately provides a unified objective-function representation that can subsequently be used by both the deterministic and robust optimization models.

The infrastructure investment cost is expressed as:
\begin{align}
\label{eq:cap_cost}
C^{\mathrm{cap}}
&=
\sum_{n\in\mathcal{N}}
\sum_{c\in\mathcal{C}^{ev}}
\kappa_c N_{n,c}
+
\sum_{d\in\mathcal{D}}
\sum_{c\in\mathcal{C}^{bus}}
\kappa_c N_{d,c},
\end{align}
where $\kappa_c$ denotes the investment cost of charger type $c$, while $N_{n,c}$ and $N_{d,c}$ represent the installed numbers of EV and eBus chargers, respectively.

For a given realization of the operational variables, electricity imported from the utility grid incurs an operating cost, whereas locally utilized PV generation provides an operational benefit. These quantities are defined as:
\begin{align}
\label{eq:grid_cost}
C^{\mathrm{grid}}
&=
\sum_{n\in\mathcal{N}}
\sum_{t\in\mathcal{T}}
\lambda \cdot g_{n,t} \cdot \Delta t, && \forall n,t,
\\
\label{eq:pv_benefit}
C^{\mathrm{pv}}
&=
\omega
\sum_{n\in\mathcal{N}}
\sum_{t\in\mathcal{T}}
q^{pv}_{n,t} \cdot \Delta t, && \forall n,t,
\end{align}
where $\lambda$ is the electricity import price, $g_{n,t}$ denotes the imported grid power, and $\omega$ represents the weighting coefficient assigned to locally self-consumed PV power.
When combined eBus and EV V2G is enabled, battery discharge incurs a throughput cost defined as:

\begin{equation}
\label{eq:v2g_cost}
\begin{split}
C^{\mathrm{v2g}}
=
\gamma^{\mathrm{v2g}}
\Big(
&\sum_{v\in\mathcal V}\sum_{t\in\mathcal T}
p^{\mathrm{ev},-}_{v,t}\cdot\Delta t
\\
&+
\sum_{b\in\mathcal B}\sum_{t\in\mathcal T}
\sum_{d\in\mathcal D}
p^{\mathrm{bus},-}_{b,t,d}\cdot\Delta t
\Big),
\quad \forall t,
\end{split}
\end{equation}
where $\gamma^{\mathrm{v2g}}$ is the battery-throughput cost coefficient. For unidirectional charging, $C^{\mathrm{v2g}}=0$.

Since the planning problem is formulated using soft operational constraints, residual violations are explicitly quantified through unmet charging energy and terminal SOC shortfall. The total violation is expressed in equivalent energy units as:
\begin{align}
\label{eq:total_violation}
\Omega
&=
\sum_{v\in\mathcal{V}}
\xi^{ev}_{v}
+
\sum_{b\in\mathcal{B}}
\xi^{bus}_{b}
\nonumber\\
&\quad+
E^{ev}
\sum_{v\in\mathcal{V}}
\sigma^{ev}_{v}
+
E^{bus}
\sum_{b\in\mathcal{B}}
\sigma^{bus}_{b}.
\end{align}
The total soft-feasibility penalty is then computed as:
\begin{equation}
\label{eq:slack_cost}
C^{\mathrm{slack}}
=
M\Omega,
\end{equation}
where $M$ is the penalty coefficient assigned to each unit of equivalent energy violation.
The first two terms of~\eqref{eq:total_violation} quantify unmet charging demand, whereas the last two terms account for residual terminal SOC deficits converted into equivalent energy using the corresponding battery capacities.

\subsection{Centralized deterministic MILP}

The centralized deterministic formulation considers the nominal charging-demand scenario and jointly optimizes infrastructure planning and operational scheduling. The model determines charger deployment,
eBus-to-depot assignments, charging/discharging schedules, battery SOC trajectories, PV self-consumption, and grid imports while satisfying
the common physical constraints presented in Section~II. 
The deterministic objective minimizes the total infrastructure planning and fleet-operating cost.
\begin{equation}
\label{eq:central_obj}
\min
\left(
C^{\mathrm{cap}}
+
C^{\mathrm{grid}}
-
C^{\mathrm{pv}}
+
\delta_{\mathrm{V2G}} \cdot C^{\mathrm{v2g}}
+
C^{\mathrm{slack}}
\right).
\end{equation}

In \eqref{eq:central_obj}, $\delta_{\mathrm{V2G}}$ is an indicator parameter equal to one for the V2G-enabled formulation and zero otherwise. The corresponding optimal objective value is denoted by $C^{\mathrm{cen}}$.
\textcolor{black}{Thus, the centralized deterministic MILP is defined by objective~\eqref{eq:central_obj} subject to the common physical constraints~\eqref{eq:bus_assignment}--\eqref{eq:nodal_hosting_det}, under the nominal demand realization and a single set of operational decision variables.}

\subsection{Two-stage robust MILP}
\label{sec:robust_formulation}

To account for uncertainty in charging-energy demand, the deterministic formulation is extended to a scenario-based two-stage robust min--max MILP. The scenario set $\mathcal{S}$ consists of optimistic, nominal, and pessimistic transportation-energy realizations with multipliers $\mu_s\in\{0.90,1.00,1.20\}$.

\subsubsection{Scenario structure}

The transportation-energy realization associated with scenario $s$ is represented by the multiplier $\mu_s$.
For EV $v$, the scenario-dependent daily energy requirement is defined as:
\begin{equation}
E^{\mathrm{req}}_{v,s}
=
\mu_s E^{\mathrm{req}}_{v},
\qquad
\forall v\in\mathcal{V},\;
s\in\mathcal{S}.
\label{eq:ev_scenario_demand}
\end{equation}
To ensure consistency between the daily energy requirement and the terminal battery condition, the EV terminal SOC target is computed as:
\begin{equation}
\overline{\mathrm{SOC}}^{\mathrm{ev}}_{v,s}
=
\mathrm{SOC}^{\mathrm{ev},0}_{v}
+
\frac{E^{\mathrm{req}}_{v,s}}
     {E^{\mathrm{ev}}},
\qquad
\forall v\in\mathcal{V},\;
s\in\mathcal{S}.
\label{eq:ev_scenario_target}
\end{equation}
For the investigated demand profiles, all values obtained from \eqref{eq:ev_scenario_target} remain within the physical EV SOC upper bound; therefore, no target clipping is required.

For eBus $b$, uncertainty is introduced through the service-energy
profile as:
\begin{equation}
e^{\mathrm{serv}}_{b,t,s}
=
\mu_s e^{\mathrm{serv}}_{b,t},
\qquad
\forall b\in\mathcal{B},\;
t\in\mathcal{T},\;
s\in\mathcal{S}.
\label{eq:bus_scenario_service}
\end{equation}
The scenario multiplier scales the complete route-segment service-energy profile without changing its temporal structure. The separate eBus charging-energy floor $E^{\mathrm{req}}_{b}$ remains nominal in every scenario.

Our robust formulation follows a two-stage structure. 
The first-stage planning variables represent long-term planning decisions shared by all scenarios and comprise the installed charger quantities and eBus-to-depot assignment:
\begin{align}
\label{eq:shared_infra}
N_{n,c}
&\in
\mathbb{Z}_{+},
&& \forall n\in\mathcal N,\;
c\in\mathcal C^{\mathrm{ev}},
\\
N_{d,c} &\in \mathbb{Z}_{+},
&& \forall d\in\mathcal D,\;
c\in\mathcal C^{\mathrm{bus}},
\\
\label{eq:shared_bus_assignment}
x_{b,d}
&\in
\{0,1\},
&& \forall b\in\mathcal B,\;
d\in\mathcal D.
\end{align}
\textcolor{black}{All remaining operational decision variables in robust formulation are scenario-indexed and second-stage recourse variables. These include EV and eBus charging/discharging powers, plug-in and charger-state variables, SOC trajectories, grid import including PV self-consumption and curtailment, and soft-feasibility slack variables. Thus, for each $s\in\mathcal{S}$, the common physical constraints in Sec.~II-C are instantiated using the corresponding scenario-dependent operational variables.}

Therefore, the hard charger-power capacity constraints \eqref{eq:bus_charge_capacity}, \eqref{eq:bus_discharge_capacity}, \eqref{eq:ev_node_charge_power}, and \eqref{eq:ev_node_discharge_power} are replaced by the scenario-specific relaxed capacity constraints whereas the charger-port occupancy constraints remain hard.
In particular, the eBus SOC dynamics in \eqref{eq:bus_soc_v2g}, the EV terminal-SOC
constraint in \eqref{eq:ev_soc_final_soft}, and the EV energy-delivery constraint in \eqref{eq:ev_energy_req_soft}
use $e^{\mathrm{serv}}_{b,t,s}$,
$\mathrm{SOC}^{\mathrm{ev}}_{v,s}$, and $E^{\mathrm{req}}_{v,s}$, respectively. The eBus energy-delivery constraint in \eqref{eq:bus_energy_req_soft} retains the nominal $E^{\mathrm{req}}_b$.

The deterministic charging-hosting envelop \eqref{eq:nodal_hosting_det} is likewise instantiated for each scenario. 
Under stressed transportation demand, the resulting net charging demand may exceed the available nodal charging-hosting capacity. The robust formulation therefore permits a scenario-specific relaxation through the nonnegative slack $\rho^{\mathrm{grid}}_{n,t,s}$:
\begin{align}
\label{eq:nodal_hosting_rob}
D^{net}_{n,t,s}
&\le
\bar P^{node}_{n}
+
\rho^{grid}_{n,t,s},
&&
\forall n\in\mathcal N,\;
t\in\mathcal T,\;
s\in\mathcal S,
\\
\label{eq:grid_slack_nonnegative}
\rho^{grid}_{n,t,s}
&\ge 0,
&&
\forall n\in\mathcal N,\;
t\in\mathcal T,\;
s\in\mathcal S.
\end{align} 
For the V2G operation, nodal discharge reduces the net power demand $D^{\mathrm{net}}_{n,t,s}$.

\subsubsection {Robust feasibility relaxation}

Under stressed scenarios, the charger-capacity constraint for all $t$ is relaxed through the nonnegative scenario-specific slack $\rho^{\mathrm{ch}}_{n,t,s}$ as:
\begin{align}
P^{ch}_{n,t,s}
&=
\sum_{v\in\mathcal V_{n,t}} p^{ev,+}_{v,t,s}
+
\mathbb{I}_{n\in\mathcal{D}}
\sum_{b\in\mathcal B} p^{bus,+}_{b,t,n,s},
\label{eq:robust_capacity_slack}\\
P^{\mathrm{ch}}_{n,t,s}
\leq\;&
\bar P^{f}\left(N_{n,f1}+4N_{n,fm}\right)
+\bar P^{s}\left(N_{n,s1}+4N_{n,sm}\right)
\nonumber\\
&+\mathbb{I}_{n\in\mathcal{D}}
\bar P^{\mathrm{bus}}
\left(N_{n,bs1}+2N_{n,bm}\right)
+\rho^{\mathrm{ch}}_{n,t,s}.
\label{eq:capacity_slack}
\end{align}
In~\eqref{eq:robust_capacity_slack}, $P^{ch}_{n,t,s}$ is the aggregate gross charging power served at node $n$ during time interval $t$ in scenario $s$.

For the V2G-enabled formulation, the aggregate discharging power is also restricted by the installed bidirectional charger capacity. The total discharging power for all $t$ at node $n$ is defined as:
\begin{equation}
\label{eq:pdis_def}
P^{dis}_{n,t,s}
=
\sum_{v\in\mathcal V_{n,t}}
p^{ev,-}_{v,t,s}
+
\mathbb{I}_{n\in\mathcal{D}}
\sum_{b\in\mathcal B}
p^{bus,-}_{b,t,n,s},
\qquad
\forall n,t,s,
\end{equation}
which is constrained by:
\begin{align}
P^{\mathrm{dis}}_{n,t,s}
\leq\;&
\bar P^{f}\left(N_{n,f1}+4N_{n,fm}\right)
+\bar P^{s}\left(N_{n,s1}+4N_{n,sm}\right)
\nonumber\\
&+\mathbb{I}_{n\in\mathcal{D}}
\bar P^{\mathrm{bus}}
\left(N_{n,bs1}+2N_{n,bm}\right)
+\rho^{\mathrm{ch}}_{n,t,s}.
\label{eq:discharge_capacity_slack}
\end{align}
\textcolor{black}{
The total violation in scenario $s$ combines unmet charging energy, terminal SOC deficits, charger-capacity slack, and nodal hosting-capacity slack:
\begin{equation}
\begin{aligned}
\Omega_s
={}&
\sum_{v\in\mathcal{V}}
\xi^{\mathrm{ev}}_{v,s}
+
\sum_{b\in\mathcal{B}}
\xi^{\mathrm{bus}}_{b,s}
\\
&+
E^{\mathrm{ev}}
\sum_{v\in\mathcal{V}}
\sigma^{\mathrm{ev}}_{v,s}
+
E^{\mathrm{bus}}
\sum_{b\in\mathcal{B}}
\sigma^{\mathrm{bus}}_{b,s}
\\
&+
\sum_{n\in\mathcal{N}}
\sum_{t\in\mathcal{T}}
\left(
\rho^{\mathrm{ch}}_{n,t,s}
+
\rho^{\mathrm{grid}}_{n,t,s}
\right) \cdot \Delta t,
\qquad
\forall s\in\mathcal{S}.
\end{aligned}
\label{eq:scenario_violation}
\end{equation}
The corresponding scenario-specific soft-feasibility penalty is:
\begin{equation}
C^{\mathrm{slack}}_s
=
M\Omega_s,
\qquad
\forall s\in\mathcal{S}.
\label{eq:scenario_slack_cost}
\end{equation}
}
Because capacity slack is explicitly penalized in the objective, its use represents a quantified relaxation of the installed-capacity requirement under stressed scenarios rather than unreported model infeasibility.

\subsubsection{Robust objective}

The robust objective minimizes the infrastructure investment together with the worst-case operating cost across all demand scenarios. Introducing the auxiliary epigraph variable $\theta$, the optimization problem is formulated as:

\begin{equation}
\label{eq:robust_obj}
\min
\left(
C^{\mathrm{cap}}
+
\theta
\right).
\end{equation}
The epigraph formulation in \eqref{eq:robust_obj} is equivalent to the following min--max optimization problem:
\begin{equation}
\label{eq:minmax_equiv}
\min
\left[
C^{\mathrm{cap}}
+
\max_{s\in\mathcal S}
C^{\mathrm{op}}_{s}
\right].
\end{equation}

The operating cost of each scenario consists of the grid-import cost, PV self-consumption benefit, soft-feasibility penalty, and, for V2G-enabled operation, battery-throughput cost. The scenario-specific quantities $C_s^{\mathrm{grid}}$, $C_s^{\mathrm{pv}}$, and $C_s^{\mathrm{v2g}}$ are obtained from their corresponding common cost expressions using scenario-indexed operational variables, while $C_s^{\mathrm{slack}}$ is defined subsequently from the scenario-wise feasibility violation. For unidirectional charging, $C_s^{\mathrm{v2g}}=0$.
Accordingly, the scenario operating cost is:
\begin{align}
\label{eq:robust_op_cost}
C^{\mathrm{op}}_{s}
&=
C^{\mathrm{grid}}_{s}
-
C^{\mathrm{pv}}_{s}
+
C^{\mathrm{slack}}_{s}
+
\delta_{\mathrm{V2G}}C^{v2g}_{s},
&&
\forall s\in\mathcal S,
\\
\label{eq:theta_constraint}
\theta
&\ge
C^{\mathrm{op}}_{s},
&&
\forall s\in\mathcal S.
\end{align}
Eq.~\eqref{eq:robust_op_cost} 
represents both the unidirectional and V2G-enabled case within a unified notation.
The epigraph constraints ensures that $\theta$ upper-bounds the operating cost in every scenario; thus, minimizing $\theta$ is equivalent to minimizing the worst-case operational expenditure.
\textcolor{black}{Accordingly, the robust MILP comprises the shared first-stage variables~\eqref{eq:shared_infra}--\eqref{eq:shared_bus_assignment}
, the second-stage recourse variables, the demand relations~\eqref{eq:ev_scenario_demand}--\eqref{eq:bus_scenario_service}, the common physical constraints, the relaxed nodal and charger-capacity constraints \eqref{eq:nodal_hosting_rob}--\eqref{eq:discharge_capacity_slack} for each $s\in\mathcal{S}$, the scenario operating-cost~\eqref{eq:robust_op_cost}, and the epigraph constraints~\eqref{eq:theta_constraint}.}

\subsection{Solution evaluation and feasibility assessment}

Soft-feasibility penalties preserve solvability under stressed conditions but are not physical expenditures. We therefore distinguish the scenario cost excluding slack penalties from the penalized objective metric as:
\begin{equation}
C^{real}_{s}
=
C^{cap}
+
C^{grid}_{s}
-
C^{pv}_{s}
+
\delta_{\mathrm{V2G}}C^{v2g}_{s},
\qquad
\forall s\in\mathcal S.
\label{eq:scenario_real_cost}
\end{equation}
The penalized scenario cost and its robust worst-case counterpart are:
\begin{align}
\label{eq:penalized_cost_s}
C^{\mathrm{pen}}_{s}
&=
C^{\mathrm{real}}_{s}
+
M\Omega_s,
&& \forall s\in\mathcal{S},\\
\label{eq:reported_robust_cost}
C^{\mathrm{rob}}
&=
\max_{s\in\mathcal{S}}
C^{\mathrm{pen}}_{s}.
\end{align}
The feasibility of the obtained solution is classified using the scenario-wise violation measure:
\begin{align}
\label{eq:hard_feasible}
\Omega_s
&=
0,
&& \forall s\in\mathcal{S}
\quad \text{(hard-feasible)},\\
\label{eq:soft_feasible}
\Omega_{\bar{s}}
&>
0,
&&
\exists \bar{s}\in\mathcal{S}
\quad \text{(soft-feasible)}.
\end{align}
Reporting $C_{s}^{\mathrm{real}}$ and $\Omega_{s,\bar{s}}$ separately ensures transparent performance reporting across all parameter sweeps.

\section{Case study setup}

\subsection{Test system and grid model}

The numerical study considers a radial 50-node distribution network comprising 40 EVs and 10 eBuses charging over a 24-hour horizon, consistent with recent distribution-network planning studies that employ medium-scale radial 50-node test systems~\cite{11267442}. Candidate eBus depots are located at nodes 1 and 49, while the slack node (node 0) is excluded from charging infrastructure placement.
The distribution grid is modeled with node-level active-power hosting limits instead of a detailed AC power-flow model. Nodal charging-hosting limits are parameterized from corresponding transformer MVA ratings using a 0.95 power factor.
PV is installed at nodes $\{1,\ldots,10,21,\ldots,40,49\}$ at 50~kW each, following a common normalized daytime profile. 

\subsection{Infrastructure, demand, and scenario parameters}

Table~\ref{tab:infrastructure_specs} summarizes the charging-infrastructure parameters, with light-duty EVSE costs benchmarked against national infrastructure studies~\cite{nrel_costs_2023}. 
Unless otherwise stated, the baseline configuration uses $M=10$~k\$/kWh, and $\omega=0$. Section~V-B explicitly varies these parameters in the sensitivity analysis. The baseline grid-import price is \$0.15/kWh.

\begin{table}[htb!]
\centering
\caption{Techno-economic specifications for charging infrastructure.}
\label{tab:infrastructure_specs}
\small
\renewcommand{\arraystretch}{1.2}
\begin{tabular*}{\columnwidth}{@{\extracolsep{\fill}} l l c c c @{}}
\toprule
\textbf{Asset} & \textbf{Topology} & \textbf{Power} & \textbf{Ports} & \textbf{Cost} \\
\textbf{Class} & & \textbf{(kW)} & & \textbf{(k\$)} \\
\midrule
AC Slow EV   & Single-port & 7   & 1 & 1.5 \\
AC Slow EV   & Multi-port  & 7   & 4 & 5.0 \\
DC Fast EV   & Single-port & 50  & 1 & 50.0 \\
DC Fast EV   & Multi-port  & 50  & 4 & 150.0 \\
DC eBus      & Single-port & 150 & 1 & 100.0 \\
DC eBus      & Multi-port  & 150 & 2 & 200.0 \\
\bottomrule
\end{tabular*}
\end{table}

EVs follow representative shopper, commuter, and delivery profiles and use 40-kWh batteries with 20\% initial SOC and 85\% charging efficiency. Their nominal charging-energy requirements are 8, 12, and 16~kWh, respectively, corresponding to terminal SOC targets of 40\%, 50\%, and 60\%.
Each eBus uses a 250-kWh battery with 15\% initial SOC, a 95\% terminal SOC target, and 90\% charging efficiency.

The robust case uses the three scenarios from Section~III-C, with $\mu_s\in\{0.90,1.00,1.20\}$ for optimistic, nominal, and pessimistic realizations, respectively. For EVs, these multipliers scale the daily charging-energy requirement and terminal SOC target. 
For the case study, depot availability follows the eBus parking schedules: buses are available for depot charging at $t\in\{0,\ldots,5,22,23\}$ and unavailable during service hours $t\in\{6,\ldots,21\}$. This schedule specifies $A^{\mathrm{dep}}_{b,d,t}$ in our numerical model. The effective route-energy coefficient $\alpha_r$ is calibrated for each eBus so that its daily route-segment energy matches the prescribed service-energy requirement.
Vehicle availability, route timing, depot availability, grid limits, electricity prices, and PV profiles are unchanged across scenarios $s\in\mathcal{S}$.

%


\section{Results and Discussion}

All reported results in this section correspond to the 50-node distribution system and use the route-segment eBus service-energy representation described in Section~IV-B. Section~V-A evaluates the effect of increasing eBus service-energy demand, whereas Section~V-B examines the effects of parameters ($M$, $\omega$), on planning.
All models are implemented in Python using PuLP \cite{mitchell2011pulp} and solved with the CBC MILP solver \cite{forrest2005cbc, lougee2003common} using a relative MIP optimality gap of 1\%.

\subsection{Impact of route-segment eBus service-energy demand}

To examine the interaction between transportation demand and infrastructure sizing, the route-segment eBus service-energy requirement is increased by up to 40\% relative to the baseline, while EV demand, network topology, and other optimization settings remain unchanged. The resulting infrastructure plans are summarized in Table~\ref{tab:segment_sensitivity}.

At the nominal demand level, both formulations install 30 chargers, but the robust solution uses a different eBus charger mix and incurs 38.4~kWh of worst-case equivalent violation. With a 10\% increase in service-energy demand, the centralized solution retains essentially the same infrastructure cost, whereas the robust solution increases its deployment to 31 chargers and eliminates the residual violation.
The robust deployment remains unchanged at the 30\% increase, before rising to 35 chargers at the 40\% increase.

Table~\ref{tab:segment_sensitivity} indicates that higher eBus service-energy demand is accommodated through both additional charger ports and changes in charger technology.
The centralized solution shifts to single-port eBus chargers at higher demand, while the robust formulation keeps a multi-port configuration over intermediate demand levels. This highlights the benefit of jointly optimizing charger technology and operational feasibility rather than sizing infrastructure from a fixed charging schedule.

\begin{table*}[t]
\centering
\small
\caption{Impact of increasing route-segment service-energy demand on charging infrastructure planning}
\label{tab:segment_sensitivity}
\renewcommand{\arraystretch}{1.2}
\begin{tabular}{c c c c c c c c c}
\toprule
\multirow{2}{*}{\textbf{\makecell{Demand\\scenario}}} & 
\multirow{2}{*}{\textbf{Method}} & 
\multicolumn{5}{c}{\textbf{Number of chargers}} & 
\textbf{Infrastructure Cost} & 
\textbf{Violation} \\
\cmidrule(lr){3-7}

& & 
\textbf{EV Slow} & 
\textbf{EV Fast} & 
\textbf{Bus Single} & 
\textbf{Bus Multi} & 
\textbf{Total} & 
\textbf{(k\$)} & 
\textbf{(kWh)} \\
\midrule
\hline

Baseline  & Centralized & 27 & 0 & 1 & 2 & 30 & 540.5 & 0.0 \\
     & Robust      & 26 & 0 & 0 & 4 & 30 & 839.0 & 38.4 \\
\midrule

Baseline + 10\% & Centralized & 27 & 0 & 1 & 2 & 30 & 540.5 & 0.0 \\
     & Robust      & 27 & 0 & 0 & 4 & 31 & 840.5 & 0.0 \\
\midrule

Baseline + 30\% & Centralized & 27 & 0 & 7 & 0 & 34 & 740.5 & 0.0 \\
     & Robust      & 27 & 0 & 0 & 4 & 31 & 840.5 & 0.0 \\
\midrule

Baseline + 40\% & Centralized & 27 & 0 & 7 & 0 & 34 & 740.5 & 0.0 \\
     & Robust      & 29 & 0 & 4 & 2 & 35 & 847.0 & 0.0 \\
\bottomrule
\end{tabular}
\end{table*}

\subsection{Sensitivity analysis of planning parameters}

\begin{figure*}[t!]
    \centering
    \begin{minipage}[t]{0.32\textwidth}
        \centering
        \begin{tikzpicture}

\begin{axis}[
    width=0.92\linewidth,
    height=0.84\linewidth,
    ybar,
    bar width=7pt,
    enlarge x limits=0.20,
    symbolic x coords={5,10,20,50},
    xtick=data,
    xlabel={$M$ (k\$/kWh)},
    ylabel={Slack penalty (k\$)},
    ymin=0,
    ymax=500,
    ytick={0,100,200,300,400,500},
    grid=major,
    grid style={dashed,gray!25},
    tick label style={font=\scriptsize},
    label style={font=\scriptsize},
    ylabel style={font=\scriptsize,yshift=-2pt},
    xlabel style={font=\scriptsize,yshift=1pt},
    legend style={
        font=\tiny,
        at={(0.5,1.02)},
        anchor=south,
        legend columns=2,
        draw=gray!60,
        fill=white,
        fill opacity=0.95,
        text opacity=1,
        inner xsep=3pt,
        inner ysep=2pt,
        row sep=1pt,
        /tikz/every even column/.append style={
            column sep=5pt
        }
    }
]

\addplot+[
    fill=blue!35,
    draw=blue!70
]
coordinates {
    (5,0.0000)
    (10,0.0000)
    (20,0.0000)
    (50,0.0000)
};
\addlegendentry{Centralized penalty}

\addplot+[
    fill=green!40,
    draw=green!60!black
]
coordinates {
    (5,135.0357)
    (10,90.0000)
    (20,180.0391)
    (50,450.0000)
};
\addlegendentry{Robust penalty}

\addlegendimage{
    legend image code/.code={
        \draw[blue,dashed,thick]
        (0cm,0cm)--(0.35cm,0cm);
        \filldraw[blue]
        (0.175cm,0cm) circle (1.2pt);
    }
}
\addlegendentry{Centralized violation}

\addlegendimage{
    legend image code/.code={
        \draw[red,thick]
        (0cm,0cm)--(0.35cm,0cm);
        \filldraw[red]
        (0.175cm,0cm) circle (1.2pt);
    }
}
\addlegendentry{Robust violation}

\end{axis}

\begin{axis}[
    width=0.92\linewidth,
    height=0.84\linewidth,
    axis y line*=right,
    axis x line=none,
    symbolic x coords={5,10,20,50},
    xtick=data,
    ylabel={Soft-feasibility violation (kWh-eqv.)},
    ymin=0,
    ymax=30,
    ytick={0,5,10,15,20,25,30},
    tick label style={font=\scriptsize},
    label style={font=\scriptsize},
    ylabel style={font=\scriptsize,yshift=3pt}
]

\addplot[
    blue,
    dashed,
    mark=*,
    thick
]
coordinates {
    (5,0.0000)
    (10,0.0000)
    (20,0.0000)
    (50,0.0000)
};

\addplot[
    red,
    solid,
    mark=*,
    thick
]
coordinates {
    (5,27.0071)
    (10,9.0000)
    (20,9.0020)
    (50,9.0000)
};


\end{axis}

\end{tikzpicture}
        \vspace{-3.5mm}
        \subcaption{$M$ sweep (Unidirectional).}
        \label{fig:m_sensitivity_no_v2g}
    \end{minipage}
    \hfill
    \begin{minipage}[t]{0.32\textwidth}
        \centering
        \raisebox{2.1mm}{%
\begin{tikzpicture}

\begin{axis}[
    width=0.92\linewidth,
    height=0.84\linewidth,
    symbolic x coords={0.00,0.02,0.05,0.10},
    xtick=data,
    enlarge x limits=0.20,
    xlabel={$\omega$},
    ylabel={Nominal grid import (kWh)},
    ymin=3500,
    ymax=4300,
    ytick={3500,3700,3900,4100,4300},
    grid=major,
    grid style={dashed,gray!25},
    tick label style={font=\scriptsize},
    xticklabel style={font=\tiny},
    label style={font=\scriptsize},
    ylabel style={font=\scriptsize,yshift=-2pt},
    xlabel style={font=\scriptsize,yshift=1pt},
    scaled y ticks=false,
    yticklabel style={
        font=\scriptsize,
        /pgf/number format/fixed,
        /pgf/number format/1000 sep={\,}
    },
    legend style={
        font=\tiny,
        at={(0.5,1.02)},
        anchor=south,
        legend columns=2,
        draw=gray!60,
        fill=white,
        fill opacity=0.95,
        text opacity=1,
        inner xsep=3pt,
        inner ysep=2pt,
        row sep=1pt,
        /tikz/every even column/.append style={
            column sep=5pt
        }
    }
]

\addplot[
    blue,
    dashed,
    mark=*,
    thick
]
coordinates {
    (0.00,3588.84)
    (0.02,3602.84)
    (0.05,3576.78)
    (0.10,3569.54)
};
\addlegendentry{Centralized grid import}

\addplot[
    red,
    solid,
    mark=*,
    thick
]
coordinates {
    (0.00,3595.84)
    (0.02,3849.02)
    (0.05,4127.36)
    (0.10,4221.02)
};
\addlegendentry{Robust grid import}

\addlegendimage{
    legend image code/.code={
        \draw[black,dash dot,thick]
        (0cm,0cm)--(0.35cm,0cm);
        \filldraw[black]
        (0.175cm,0cm) rectangle
        ++(0.055cm,0.055cm);
    }
}
\addlegendentry{Robust worst-case violation}

\addlegendimage{empty legend}
\addlegendentry{\phantom{Infrastructure cost}}

\end{axis}

\begin{axis}[
    width=0.92\linewidth,
    height=0.84\linewidth,
    axis y line*=right,
    axis x line=none,
    symbolic x coords={0.00,0.02,0.05,0.10},
    xtick=data,
    enlarge x limits=0.20,
    ylabel={Worst-case violation (kWh-eqv.)},
    ymin=0,
    ymax=10,
    ytick={0,2,4,6,8,10},
    tick label style={font=\scriptsize},
    label style={font=\scriptsize},
    ylabel style={font=\scriptsize,yshift=3pt}
]

\addplot[
    black,
    dash dot,
    mark=square*,
    thick
]
coordinates {
    (0.00,9.0000)
    (0.02,8.0606)
    (0.05,5.7095)
    (0.10,1.6689)
};


\end{axis}

\end{tikzpicture}%
}
        \vspace{-3.5mm}
        \subcaption{$\omega$ sweep (Unidirectional).}
        \label{fig:gridImport_no_v2g}
    \end{minipage}
    \hfill
    \begin{minipage}[t]{0.32\textwidth}
        \centering
        \begin{tikzpicture}

\begin{axis}[
    width=0.92\linewidth,
    height=0.84\linewidth,
    ybar,
    bar width=7pt,
    enlarge x limits=0.20,
    symbolic x coords={0.00,0.02,0.05,0.10},
    xtick=data,
    xlabel={$\omega$},
    ylabel={Total installed chargers},
    ymin=0,
    ymax=48,
    ytick={0,10,20,30,40},
    grid=major,
    grid style={dashed,gray!25},
    tick label style={font=\scriptsize},
    xticklabel style={font=\tiny},
    label style={font=\scriptsize},
    ylabel style={font=\scriptsize,yshift=-2pt},
    xlabel style={font=\scriptsize,yshift=1pt},
    legend style={
        font=\tiny,
        at={(0.5,1.02)},
        anchor=south,
        legend columns=2,
        draw=gray!60,
        fill=white,
        fill opacity=0.95,
        text opacity=1,
        inner xsep=3pt,
        inner ysep=2pt,
        row sep=1pt,
        /tikz/every even column/.append style={
            column sep=5pt
        }
    }
]

\addplot+[
    fill=blue!35,
    draw=blue!70
]
coordinates {
    (0.00,30)
    (0.02,31)
    (0.05,36)
    (0.10,37)
};
\addlegendentry{Centralized chargers}

\addplot+[
    fill=green!40,
    draw=green!60!black
]
coordinates {
    (0.00,31)
    (0.02,33)
    (0.05,38)
    (0.10,41)
};
\addlegendentry{Robust chargers}

\addlegendimage{
    legend image code/.code={
        \draw[blue,dashed,thick]
        (0cm,0cm)--(0.35cm,0cm);
        \filldraw[blue]
        (0.175cm,0cm) circle (1.2pt);
    }
}
\addlegendentry{Centralized infrastructure cost}

\addlegendimage{
    legend image code/.code={
        \draw[red,thick]
        (0cm,0cm)--(0.35cm,0cm);
        \filldraw[red]
        (0.175cm,0cm) circle (1.2pt);
    }
}
\addlegendentry{Robust infrastructure cost}

\end{axis}

\begin{axis}[
    width=0.92\linewidth,
    height=0.84\linewidth,
    axis y line*=right,
    axis x line=none,
    symbolic x coords={0.00,0.02,0.05,0.10},
    xtick=data,
    ylabel={Infrastructure cost (k\$)},
    ymin=530,
    ymax=670,
    ytick={540,570,600,630,660},
    tick label style={font=\scriptsize},
    label style={font=\scriptsize},
    ylabel style={font=\scriptsize,yshift=3pt}
]

\addplot[
    blue,
    dashed,
    mark=*,
    thick
]
coordinates {
    (0.00,540.5)
    (0.02,542.0)
    (0.05,549.5)
    (0.10,551.0)
};

\addplot[
    red,
    solid,
    mark=*,
    thick
]
coordinates {
    (0.00,642.0)
    (0.02,642.0)
    (0.05,649.5)
    (0.10,654.0)
};


\end{axis}

\end{tikzpicture}
        \vspace{-3.5mm}
        \subcaption{$\omega$ sweep (Unidirectional).}
        \label{fig:omega_sensitivity_no_v2g}
    \end{minipage}
    
    \vspace{1mm}
    
    \begin{minipage}[t]{0.32\textwidth}
        \centering
        \begin{tikzpicture}

\begin{axis}[
    width=0.92\linewidth,
    height=0.84\linewidth,
    ybar,
    bar width=7pt,
    enlarge x limits=0.20,
    symbolic x coords={5,10,20,50},
    xtick=data,
    xlabel={$M$ (k\$/kWh)},
    ylabel={Slack penalty (k\$)},
    ymin=0,
    ymax=500,
    ytick={0,100,200,300,400,500},
    grid=major,
    grid style={dashed,gray!25},
    tick label style={font=\scriptsize},
    label style={font=\scriptsize},
    ylabel style={font=\scriptsize,yshift=-2pt},
    xlabel style={font=\scriptsize,yshift=1pt},
    legend style={
        font=\tiny,
        at={(0.5,1.02)},
        anchor=south,
        legend columns=2,
        draw=gray!60,
        fill=white,
        fill opacity=0.95,
        text opacity=1,
        inner xsep=3pt,
        inner ysep=2pt,
        row sep=1pt,
        /tikz/every even column/.append style={
            column sep=5pt
        }
    }
]

\addplot+[
    fill=blue!35,
    draw=blue!70
]
coordinates {
    (5,0.0000)
    (10,0.0000)
    (20,0.0000)
    (50,0.0000)
};
\addlegendentry{Centralized penalty}

\addplot+[
    fill=green!40,
    draw=green!60!black
]
coordinates {
    (5,135.0353)
    (10,90.0403)
    (20,0.0000)
    (50,0.0000)
};
\addlegendentry{Robust penalty}

\addlegendimage{
    legend image code/.code={
        \draw[blue,dashed,thick]
        (0cm,0cm)--(0.35cm,0cm);
        \filldraw[blue]
        (0.175cm,0cm) circle (1.2pt);
    }
}
\addlegendentry{Centralized violation}

\addlegendimage{
    legend image code/.code={
        \draw[red,thick]
        (0cm,0cm)--(0.35cm,0cm);
        \filldraw[red]
        (0.175cm,0cm) circle (1.2pt);
    }
}
\addlegendentry{Robust violation}

\end{axis}

\begin{axis}[
    width=0.92\linewidth,
    height=0.84\linewidth,
    axis y line*=right,
    axis x line=none,
    symbolic x coords={5,10,20,50},
    xtick=data,
    ylabel={Soft-feasibility violation (kWh-eqv.)},
    ymin=0,
    ymax=30,
    ytick={0,5,10,15,20,25,30},
    tick label style={font=\scriptsize},
    label style={font=\scriptsize},
    ylabel style={font=\scriptsize,yshift=3pt}
]

\addplot[
    blue,
    dashed,
    mark=*,
    thick
]
coordinates {
    (5,0.0000)
    (10,0.0000)
    (20,0.0000)
    (50,0.0000)
};

\addplot[
    red,
    solid,
    mark=*,
    thick
]
coordinates {
    (5,27.0071)
    (10,9.0040)
    (20,0.0000)
    (50,0.0000)
};


\end{axis}

\end{tikzpicture}
        \vspace{-3.5mm}
        \subcaption{$M$ sweep (Bidirectional V2G).}
        \label{fig:m_sensitivity_v2g}
    \end{minipage}
    \hfill
    \begin{minipage}[t]{0.32\textwidth}
        \centering
        \raisebox{2.3mm}{%
\begin{tikzpicture}

\begin{axis}[
    width=0.92\linewidth,
    height=0.84\linewidth,
    symbolic x coords={0.00,0.02,0.05,0.10},
    xtick=data,
    enlarge x limits=0.20,
    xlabel={$\omega$},
    ylabel={Nominal grid import (kWh)},
    ymin=3500,
    ymax=4050,
    ytick={3500,3700,3900,4050},
    grid=major,
    grid style={dashed,gray!25},
    tick label style={font=\scriptsize},
    xticklabel style={font=\tiny},
    label style={font=\scriptsize},
    ylabel style={font=\scriptsize,yshift=-2pt},
    xlabel style={font=\scriptsize,yshift=1pt},
    scaled y ticks=false,
    yticklabel style={
        font=\scriptsize,
        /pgf/number format/fixed,
        /pgf/number format/1000 sep={\,}
    },
    legend style={
        font=\tiny,
        at={(0.5,1.02)},
        anchor=south,
        legend columns=2,
        legend cell align={left},
        draw=gray!60,
        fill=white,
        fill opacity=0.95,
        text opacity=1,
        inner xsep=4pt,
        inner ysep=2pt,
        row sep=1pt,
        /tikz/every even column/.append style={
            column sep=7pt
        }
    }
]

\addplot[
    blue,
    dashed,
    mark=*,
    thick
]
coordinates {
    (0.00,3593.660)
    (0.02,3584.010)
    (0.05,3571.950)
    (0.10,3569.540)
};
\addlegendentry{Centralized grid import}

\addplot[
    red,
    solid,
    mark=*,
    thick
]
coordinates {
    (0.00,3588.485)
    (0.02,3822.999) 
    (0.05,3963.660)
    (0.10,3810.664)
};
\addlegendentry{Robust grid import}

\addlegendimage{
    black,
    dash dot,
    mark=square*,
    thick
}
\addlegendentry{Robust worst-case violation}

\end{axis}

\begin{axis}[
    width=0.92\linewidth,
    height=0.84\linewidth,
    axis y line*=right,
    axis x line=none,
    symbolic x coords={0.00,0.02,0.05,0.10},
    xtick=data,
    enlarge x limits=0.20,
    ylabel={Worst-case violation (kWh-eqv.)},
    ymin=0,
    ymax=22,
    ytick={0,5,10,15,20},
    tick label style={font=\scriptsize},
    label style={font=\scriptsize},
    ylabel style={font=\scriptsize,yshift=3pt}
]

\addplot[
    black,
    dash dot,
    mark=square*,
    thick
]
coordinates {
    (0.00,9.0057)
    (0.02,9.0000)
    (0.05,9.0000)
    (0.10,9.0000)
};


\end{axis}

\end{tikzpicture}%
}
        \vspace{-3.5mm}
        \subcaption{$\omega$ sweep (Bidirectional V2G).}
        \label{fig:gridImport_v2g}
    \end{minipage}
    \hfill
    \begin{minipage}[t]{0.32\textwidth}
        \centering
        \begin{tikzpicture}

\begin{axis}[
    width=0.92\linewidth,
    height=0.84\linewidth,
    ybar,
    bar width=7pt,
    enlarge x limits=0.20,
    symbolic x coords={0.00,0.02,0.05,0.10},
    xtick=data,
    xlabel={$\omega$},
    ylabel={Total installed chargers},
    ymin=0,
    ymax=48,
    ytick={0,10,20,30,40},
    grid=major,
    grid style={dashed,gray!25},
    tick label style={font=\scriptsize},
    xticklabel style={font=\tiny},
    label style={font=\scriptsize},
    ylabel style={font=\scriptsize,yshift=-2pt},
    xlabel style={font=\scriptsize,yshift=1pt},
    legend style={
        font=\tiny,
        at={(0.5,1.02)},
        anchor=south,
        legend columns=2,
        draw=gray!60,
        fill=white,
        fill opacity=0.95,
        text opacity=1,
        inner xsep=3pt,
        inner ysep=2pt,
        row sep=1pt,
        /tikz/every even column/.append style={
            column sep=5pt
        }
    }
]

\addplot+[
    fill=blue!35,
    draw=blue!70
]
coordinates {
    (0.00,30)
    (0.02,31)
    (0.05,37)
    (0.10,37)
};
\addlegendentry{Centralized chargers}

\addplot+[
    fill=green!40,
    draw=green!60!black
]
coordinates {
    (0.00,36)
    (0.02,31)
    (0.05,41)
    (0.10,41)
};
\addlegendentry{Robust chargers}

\addlegendimage{
    legend image code/.code={
        \draw[blue,dashed,thick]
        (0cm,0cm)--(0.35cm,0cm);
        \filldraw[blue]
        (0.175cm,0cm) circle (1.2pt);
    }
}
\addlegendentry{Centralized infrastructure cost}

\addlegendimage{
    legend image code/.code={
        \draw[red,thick]
        (0cm,0cm)--(0.35cm,0cm);
        \filldraw[red]
        (0.175cm,0cm) circle (1.2pt);
    }
}
\addlegendentry{Robust infrastructure cost}

\end{axis}

\begin{axis}[
    width=0.92\linewidth,
    height=0.84\linewidth,
    axis y line*=right,
    axis x line=none,
    symbolic x coords={0.00,0.02,0.05,0.10},
    xtick=data,
    ylabel={Infrastructure cost (k\$)},
    ymin=520,
    ymax=760,
    ytick={540,600,660,720},
    tick label style={font=\scriptsize},
    label style={font=\scriptsize},
    ylabel style={font=\scriptsize,yshift=3pt}
]

\addplot[
    blue,
    dashed,
    mark=*,
    thick
]
coordinates {
    (0.00,540.5)
    (0.02,542.0)
    (0.05,551.0)
    (0.10,551.0)
};

\addplot[
    red,
    solid,
    mark=*,
    thick
]
coordinates {
    (0.00,655.5)
    (0.02,640.5)
    (0.05,652.5) 
    (0.10,652.5) 
};


\end{axis}

\end{tikzpicture}
        \vspace{-3.5mm}
        \subcaption{$\omega$ sweep (Bidirectional V2G).}
        \label{fig:omega_sensitivity_v2g}
    \end{minipage}
    \caption{%
   Sensitivity to $M$ and $\omega$. For the robust solution, grid imports are shown for the nominal scenario; robust violation and penalty reflect the worst case across scenarios. Infrastructure outcomes are scenario-independent.}
    \label{fig:with_no_v2g_sensitivity_combined}
\end{figure*}
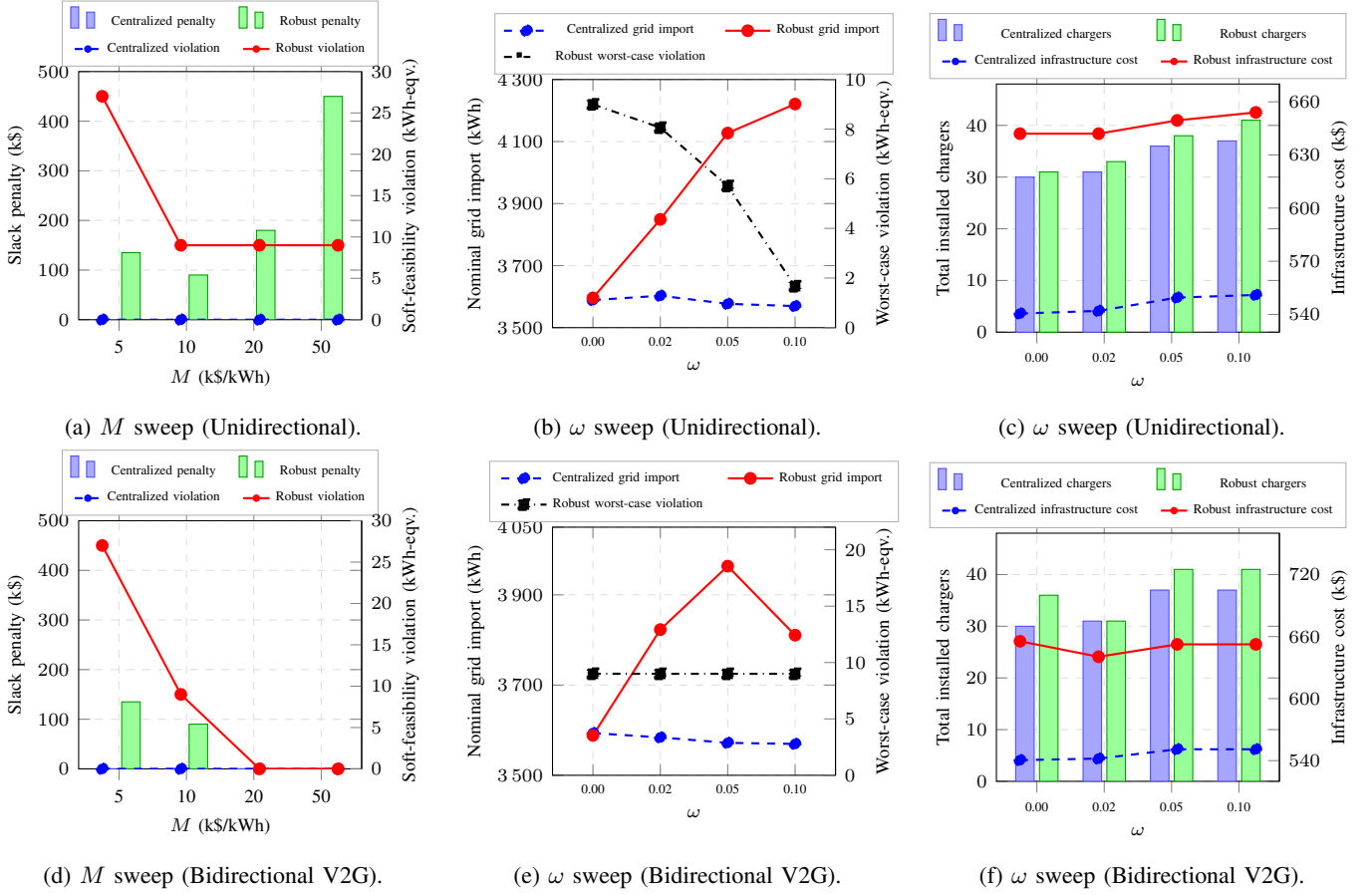

Fig.~\ref{fig:with_no_v2g_sensitivity_combined} evaluates performance of proposed framework under varying penalty parameters $M$ and PV-weighting factors $\omega$ for unidirectional and V2G operation. Centralized models optimize nominal demand, while robust models optimize scenario-dependent recourse over optimistic, nominal, and pessimistic demand realizations.
As observed, centralized solutions remain strictly violation-free across all settings. 

For robust formulations, increasing $M$ from $5$ to $10~\mathrm{k\$/kWh}$ reduces worst-case violations from $\approx 27$ to $9~\mathrm{kWh\text{-}eq.}$ under both modes. With unidirectional charging, violations stall at $\approx 9~\mathrm{kWh\text{-}eq.}$ for $M \ge 10~\mathrm{k\$/kWh}$, indicating that increasing the penalty alone does not eliminate the
residual shortfall (see Fig.~\ref{fig:m_sensitivity_no_v2g}). 
In contrast, V2G flexibility achieves 
hard feasibility ($0~\mathrm{kWh\text{-}eq.}$ violation) for $M \ge 20~\mathrm{k\$/kWh}$ (Fig.~\ref{fig:m_sensitivity_v2g}). 
This difference indicates that the additional V2G charging flexibility can absorb the stressed transportation-energy requirement without residual soft-constraint violation at sufficiently large $M$.

Fig.~\ref{fig:gridImport_no_v2g} shows that increasing $\omega$ under unidirectional mode reduces robust worst-case violations from $\approx 9$ to $<2~\mathrm{kWh\text{-}eq.}$ at the expense of higher nominal grid import. Under V2G mode, nominal grid import exhibits non-monotonic behavior while worst-case violations nearly at $9~\mathrm{kWh\text{-}eq.}$ (Fig.~\ref{fig:gridImport_v2g}), 
indicating that, for the tested
V2G configuration, changes in $\omega$ primarily affect the operating schedule and grid-import profile without materially changing the residual worst-case violation.

Infrastructure responses to $\omega$ are shown in Figs.~\ref{fig:omega_sensitivity_no_v2g} and \ref{fig:omega_sensitivity_v2g}. Under unidirectional mode, increasing $\omega$ expands centralized and robust deployments from 30 to 37 and 31 to 41 chargers, respectively. Under robust V2G, deployment varies non-monotonically ($36$, $31$, $41$, and $41$ chargers at $\omega = 0, 0.02, 0.05, 0.10$, costing $655.5$, $640.5$, $652.5$, and $652.5~\mathrm{k\$}$). This non-monotonicity is driven by discrete selection shifts among heterogeneous charger types with distinct power ratings, port counts, and unit costs.
Thus, changes in $\omega$ may alter the preferred charger mix rather than produce a monotonic change in total infrastructure requirements.

\subsection{Computational performance and scalability}

For the 50-node system, scenario expansion and V2G increase the MILP size. The robust unidirectional formulation contains approximately 32,000 variables and 46,800 constraints, increasing to about 45,000 variables and 67,500 constraints with V2G because of the additional charge and discharge logic.

Unidirectional cases solve within 8~s for the centralized model and 44~s for the robust model. Under baseline settings, V2G runtimes range from 40--121~s for the centralized formulation and 125--446~s for the robust formulation. The longest sensitivity runs are 2,602~s for the centralized case at $M=20$ and 1,883~s for the robust case at $M=10$. These solution times are suitable for offline infrastructure-planning studies.

\section{Conclusion}

This paper formulated a joint deterministic and two-stage robust MILP for EV and eBus charging-infrastructure planning with PV self-consumption and explicit soft-feasibility diagnostics. The numerical study shows that route-segment eBus demand affects not only installed capacity but also charger technology: moderate increases can be accommodated through reconfiguration, whereas larger increases trigger infrastructure expansion. The feasibility penalty governs the trade-off between investment and residual shortfall, while V2G can recover hard feasibility when unidirectional charging cannot. PV weighting further changes grid import and can produce non-monotonic investment responses because charger choices are discrete.

Our findings support treating heterogeneous fleets, scenario-adaptive operation, and nodal charging limits within a unified planning framework. The present aggregate nodal hosting limit representation excludes explicit time-varying conventional non-transport demand; future work will incorporate this effect together with dynamic electricity prices.

{
\small
\bibliographystyle{IEEEtran}
\bibliography{./references.bib}
}

\end{document}